\documentclass[aip,amsmath,amssymb,reprint]{revtex4-1}

\usepackage{bm}
\usepackage{braket}
\usepackage{graphicx}
\usepackage{hyperref}
\usepackage{mathptmx}
\usepackage{multirow}
\usepackage{xcolor}

\usepackage[utf8]{inputenc}
\usepackage[T1]{fontenc}

\newcommand{\z}{{\textcolor{white}{1}}}

\begin{document}

\title{Accurate Frozen Core Approximation for All-Electron Density-Functional Theory}

\author{Victor Wen-zhe Yu}
\affiliation{Department of Mechanical Engineering and Materials Science, Duke University, Durham, NC 27708, USA}

\author{Jonathan Moussa}
\affiliation{Molecular Sciences Software Institute, Blacksburg, VA 24060, USA}

\author{Volker Blum}
\email[Corresponding author, e-mail:~]{volker.blum@duke.edu}
\affiliation{Department of Mechanical Engineering and Materials Science, Duke University, Durham, NC 27708, USA}

\date{\today}

\begin{abstract}
We implement and benchmark the frozen core approximation, a technique commonly adopted in electronic structure theory to reduce the computational cost by means of mathematically fixing the chemically inactive core electron states. The accuracy and efficiency of this approach are well controlled by a single parameter, the number of frozen orbitals. Explicit corrections for the frozen core orbitals and the unfrozen valence orbitals are introduced, safeguarding against seemingly minor numerical deviations from the assumed orthonormality conditions of the basis functions. A speedup of over two-fold can be achieved for the diagonalization step in all-electron density-functional theory simulations containing heavy elements, without any accuracy degradation in terms of the electron density, total energy, and atomic forces. This is demonstrated in a benchmark study covering 103 materials across the periodic table, and a large-scale simulation of CsPbBr$_3$ with 2,560 atoms. Our study provides a rigorous benchmark of the precision of the frozen core approximation (sub-meV per atom for frozen core orbitals below $-200$ eV) for a wide range of test cases and for chemical elements ranging from Li to Po. The algorithms discussed here are implemented in the open-source Electronic Structure Infrastructure software package.
\end{abstract}

\maketitle

\section{Introduction}
\label{sec:intro}
Kohn-Sham density-functional theory (KS-DFT)~\cite{dft_hohenberg_1964,dft_kohn_1965} is a powerful tool to computationally predict the properties of molecules and materials. Instead of directly solving a many-electron problem, KS-DFT deals with an auxiliary set of single-electron problems, known as the Kohn-Sham equations
\begin{equation}
\label{eq:ks}
\hat{h}^\text{KS} \psi_l = \epsilon_l \psi_l .
\end{equation}

\noindent $\hat{h}^\text{KS}$ denotes the Kohn-Sham Hamiltonian. $\psi_l$ and $\epsilon_l$ are the Kohn-Sham orbitals and their eigenenergies. In practical implementations of KS-DFT, $N_\text{basis}$ basis functions $\phi_i$ are employed to expand the Kohn-Sham orbitals,
\begin{equation}
\label{eq:basis}
\psi_l = \sum_{i=1}^{N_\text{basis}} c_{il} \phi_i ,
\end{equation}

\noindent where $c_{il}$ is the expansion coefficient. This converts the Kohn-Sham equations into an algebraic eigenproblem
\begin{equation}
\label{eq:gevp}
\boldsymbol{H} \boldsymbol{C} = \boldsymbol{S} \boldsymbol{C} \boldsymbol{\Sigma} .
\end{equation}

\noindent The elements of the Hamiltonian matrix $\boldsymbol{H}$ and the overlap matrix $\boldsymbol{S}$ are computed by integrations:
\begin{subequations}
\label{eq:ham_integ}
\begin{align}
H_{(i,j)} & = \int d^3 \boldsymbol{r} [\phi_i^*(\boldsymbol{r}) \hat{h}^\text{KS} (\boldsymbol{r}) \phi_j(\boldsymbol{r})] , \\
S_{(i,j)} & = \int d^3 \boldsymbol{r} [\phi_i^*(\boldsymbol{r}) \phi_j(\boldsymbol{r})] .
\end{align}
\end{subequations}

\noindent The Kohn-Sham orbitals $\psi_l$ and their occupation numbers $f_l$ are obtained from the eigenvalues $\boldsymbol{\Sigma}$ and eigenvectors $\boldsymbol{C}$ of equation~\ref{eq:gevp}. Then the electron density $\boldsymbol{n}$ can be constructed by
\begin{equation}
\label{eq:density_from_orbital}
\boldsymbol{n}(\boldsymbol{r}) = \sum_{l} f_l \vert \psi_l(\boldsymbol{r}) \vert ^2 .
\end{equation}

\noindent Alternatively, it is possible to directly construct $\boldsymbol{n}$ from the $\boldsymbol{H}$ and $\boldsymbol{S}$ matrices without explicitly solving equation~\ref{eq:gevp}. Since $\boldsymbol{n}$ depends on $\boldsymbol{H}$, and $\boldsymbol{H}$ in turn depends on $\boldsymbol{n}$, the Kohn-Sham equations are actually a non-linear problem, and they are usually solved iteratively by a self-consistent field (SCF) approach.

The choice of the basis set is the most fundamental decision in the design of a KS-DFT code. Established choices include plane waves~\cite{pw_martin_2004,vasp_kresse_1996,castep_clark_2005,abinit_gonze_2020,qe_giannozzi_2020}, (linearized) augmented plane waves with local orbitals~\cite{lapw_singh_1994,fleur_blugel_1987,exciting_gulans_2014,wien2k_blaha_2020}, Gaussian type orbitals~\cite{gto_szabo_1989,qchem_shao_2015,gaussian_frisch_2016,cp2k_kuhne_2020,nwchem_apra_2020,psi4_smith_2020,turbomole_balasubramani_2020}, Slater type orbitals~\cite{sto_slater_1930,adf_velde_2001}, numeric atom-centered orbitals~\cite{dmol_delley_1990,fhiaims_blum_2009,quantumatk_smidstrup_2019,siesta_garcia_2020}, and many others. Regardless of the form of the basis functions, the size of the basis set, $N_\text{basis}$, generally increases with the number of electrons in the system being simulated, rendering simulations of systems containing heavy elements computationally demanding. For a broad range of chemical and physical phenomena, it suffices to study the valence and perhaps shallow ``semi-core'' electrons that actively participate in the formation of chemical bonds. Several different kinds of computational methods take advantage of this fact to solve the electronic structure problem in the space of the valence electrons. The solution of the reduced problem can match the solution of the full problem involving all the electrons, as long as the target property does not explicitly depend on the deeper, chemically essentially inert core electrons.

In the frozen core approximation~\cite{frozen_fink_1972,frozen_baerends_1973,frozen_sachs_1975,frozen_pettersson_1982,frozen_matsuoka_1992,fplo_koepernik_1999}, the orbitals corresponding to the core electrons remain fixed throughout the calculation. Correspondingly, the minimization of the KS-DFT energy functional is carried out in the variational space of the valence electrons. The orthonormality between all Kohn-Sham orbitals must be ensured, i.e., the core orbitals should be orthogonal to one another, and the valence orbitals should be orthogonal to the core orbitals. This can be enforced by orthonormalizing the valence orbitals to the core orbitals, using e.g. the Gram-Schmidt process, or by directly making the valence basis functions orthogonal to the core basis functions. In the latter case, the valence basis functions are orthonormalized to the core basis functions by Gram-Schmidt or by mixing in a linear combination of the core (or core-like) basis functions~\cite{adf_velde_2001,frozen_baerends_1973,frozen_matsuoka_1992}. These new valence basis functions then enter the variational treatment of the valence Kohn-Sham orbitals.

The frozen core approximation forms the theoretical foundation of the pseudopotential method, a corner stone of KS-DFT codes employing plane waves as basis functions. Expanding the fast oscillating core orbitals would require a prohibitively large number of plane waves, which is circumvented by using various flavors of pseudopotentials~\cite{pseudo_hellmann_1935,pseudo_gombas_1935,paw_slater_1937,pseudo_phillips_1959,pseudo_austin_1962,pseudo_hamann_1979,pseudo_vanderbilt_1990} or by the projector augmented wave method~\cite{paw_blochl_1994,paw_kresse_1999}. Encapsulating the effects of the core electrons, possibly with relativistic effects for heavy elements, pseudopotentials enter the effective Hamiltonian defined for the valence electrons as a fixed term throughout a computation. Pseudopotentials and effective core potentials~\cite{ecp_hay_1985a,ecp_hay_1985b,ecp_wadt_1985} can also be used with localized basis functions, e.g., Gaussian type orbitals~\cite{qchem_shao_2015,gaussian_frisch_2016,cp2k_kuhne_2020,nwchem_apra_2020,turbomole_balasubramani_2020} and numeric atom-centered orbitals~\cite{siesta_garcia_2020}, as a means of reducing the computational complexity. A comprehensive benchmark study for cohesive properties of solids carried out by the DFT community, the ``Delta test''~\cite{delta_lejaeghere_2016,delta_web} shows that after detailed parameterization, codes based on pseudopotentials can come close to the precision of all-electron codes. However, one downside of employing ``pseudoized'' approaches to the potential or orbitals is that the respective parameterizations must be defined and fixed \textit{a priori}, i.e., prior to executing a given calculation. Once decided, the pseudoization is typically impossible to relax within a given calculation, unless more involved techniques are used, such as the relaxed core projector augmented wave method featuring an on-the-fly repseudoization~\cite{relax_marsman_2006}. Additionally, facile access to the core orbitals, for applications such as core level spectroscopies, is typically lost in such calculations and specialized approaches are required to restore it within a pseudoized framework~\cite{pseudo_vanderbilt_1990,paw_blochl_1994,hyperfine_vandewalle_1993,nmr_pickard_2001,nmr_yates_2007,core_gao_2009}.

In this regard, applying the frozen core approximation without pseudoization has the benefit that it retains computational access to the localized, fixed core orbitals throughout the calculation. This approach has been employed in formal all-electron codes with e.g., Slater type orbitals~\cite{adf_velde_2001}, linearized augmented plan waves~\cite{fleur_blugel_1987,exciting_gulans_2014}, and linearized muffin-tin orbitals~\cite{questaal_pashov_2020}. In principle, this method can be extended to seamlessly move from an initial frozen-core treatment to the full all-electron result within a single calculation. In this paper, we demonstrate a highly accurate frozen-core approximation within the diagonalization (solution of equation~\ref{eq:gevp}) stage of all-electron KS-DFT calculations, which, in the limit of large systems, becomes the computationally dominant step of such calculations. As the computational cost of a dense eigensolver scales cubically with respect to the size of the problem, a 25\% reduction in the problem size would almost halve the cost, alleviating a computational burden in large-scale simulations containing heavy elements. The orthonormality of the Kohn-Sham orbitals is ensured even in the presence of numerical errors in the Hamiltonian and overlap integrations (equation~\ref{eq:ham_integ}). A considerable saving in the computational cost can be achieved, with physical observables, such as the total energy and atomic forces, matching the accuracy of all-electron computations to a high degree.

Our present work is restricted to the diagonalization stage of KS-DFT. Other steps of the calculation, such as the electron density update, could trivially take advantage of the results without making any further approximations, leading to significant further potential computational savings, but those steps are not the focus of this work. The emphasis of this paper is on (1) rigorously exploring the accuracy that can be retained by enforcing the frozen core approximation in an otherwise high-precision, all-electron calculation, as well as (2) quantifying the computational savings that can arise within the computationally dominant diagonalization step.

The rest of the paper is organized in four sections. Section~\ref{sec:method} briefly reviews the frozen core approximation as described by Koepernik and Eschrig in 1999~\cite{fplo_koepernik_1999} (referred to as FC99), then outlines two modifications that improve the accuracy and robustness of the method. These approaches are referred to as FC99+C and FC99+C+V. ``C'' denotes a modification to account for small numerical deviations from strict orthonormality between the frozen core orbitals. ``V'' denotes a correction that, in its essence, discards an approximation made in the FC99 method and therefore recovers the exact form of the frozen core approximation. The accuracy and computational efficiency of the FC99+C method and especially of the favored FC99+C+V method are thoroughly assessed in section~\ref{sec:accuracy} and section~\ref{sec:performance}, respectively, by a benchmark set of 103 materials covering chemical species across the periodic table and a large-scale KS-DFT simulation of CsPbBr$_3$ containing 2,560 atoms. When applied to core orbitals with energies of below $-200$ eV, the FC99+C+V method is found to be completely robust and numerically accurate to better than 1~meV/atom for total energies in our tests. Although the general validity of the frozen core approach for deep core levels is widely established in the literature, the present benchmark is the broadest and most rigorous demonstration of this degree of numerical precision for the frozen core approach known to the authors. Our conclusions are given in section~\ref{sec:summary}.

\section{Frozen Core Approximation}
\label{sec:method}
\subsection{The FC99 Method}
\label{sec:fc99}
The frozen core approximation takes the fixed, strongly localized core atomic orbitals as the core Kohn-Sham orbitals. As the latter must be orthonormalized to each other, the core atomic orbitals must already be orthonormalized. The FC99 method uses the Cholesky factorization of the overlap matrix to project the full Hamiltonian to a subspace spanned by the valence atomic orbitals. The valence Kohn-Sham orbitals are then determined variationally. The Hamiltonian and overlap matrices are partitioned into four blocks, namely the core-core (cc) block, the valence-core (vc) block, the core-valence (cv) block, and the valence-valence (vv) block, i.e.,
\begin{subequations}
\label{eq:ham_blocks}
\begin{align}
\boldsymbol{H} & =
\begin{bmatrix}
\boldsymbol{H}_\text{cc} & \boldsymbol{H}_\text{cv} \\
\boldsymbol{H}_\text{vc} & \boldsymbol{H}_\text{vv}
\end{bmatrix} , \\
\boldsymbol{S} & =
\begin{bmatrix}
\boldsymbol{S}_\text{cc} & \boldsymbol{S}_\text{cv} \\
\boldsymbol{S}_\text{vc} & \boldsymbol{S}_\text{vv}
\end{bmatrix} .
\end{align}
\end{subequations}

\noindent As the core atomic orbitals are orthonormalized, (i) $\boldsymbol{S}_\text{cc}$ is an identity matrix, and (ii) $\boldsymbol{H}_\text{cc}$ is a diagonal matrix. Because of (i), the Cholesky factorization of $\boldsymbol{S}$ takes the form
\begin{equation}
\label{eq:cholesky}
\boldsymbol{S} =
\begin{bmatrix}
\boldsymbol{I} & \boldsymbol{0} \\
\boldsymbol{S}_\text{vc} & \boldsymbol{L}_\text{vv}
\end{bmatrix}
\begin{bmatrix}
\boldsymbol{I} & \boldsymbol{S}_\text{cv} \\
\boldsymbol{0} & \boldsymbol{L}_{\text{vv}}^*
\end{bmatrix} = \boldsymbol{L} \boldsymbol{L}^* ,
\end{equation}

\noindent with
\begin{equation}
\label{eq:vv_cholesky}
\boldsymbol{L}_\text{vv} \boldsymbol{L}_{\text{vv}}^* = \boldsymbol{S}_\text{vv} - \boldsymbol{S}_\text{vc} \boldsymbol{S}_\text{cv} .
\end{equation}

\noindent Using equation~\ref{eq:cholesky}, equation~\ref{eq:gevp} can be transformed to a standard eigenproblem
\begin{subequations}
\label{eq:evp}
\begin{align}
\boldsymbol{\tilde{H}} \boldsymbol{\tilde{C}} & = \boldsymbol{\tilde{C}} \boldsymbol{\Sigma} , \label{eq:evp_matrix} \\
\boldsymbol{\tilde{H}} & = \boldsymbol{L}^{-1} \boldsymbol{H} (\boldsymbol{L}^*)^{-1} , \label{eq:evp_ham} \\
\boldsymbol{\tilde{C}} & = \boldsymbol{L}^* \boldsymbol{C} .
\end{align}
\end{subequations}

\noindent Inserting equations~\ref{eq:ham_blocks} and \ref{eq:cholesky} into equation~\ref{eq:evp_ham} yields
\begin{subequations}
\label{eq:evp_ham_blocks}
\begin{align}
\boldsymbol{\tilde{H}}_\text{cc} & = \boldsymbol{H}_\text{cc} , \\
\boldsymbol{\tilde{H}}_\text{vc} & = \boldsymbol{L}_{\text{vv}}^{-1} (\boldsymbol{H}_\text{vc} - \boldsymbol{S}_\text{vc} \boldsymbol{H}_\text{cc}) , \\
\boldsymbol{\tilde{H}}_\text{cv} & = (\boldsymbol{H}_\text{cv} - \boldsymbol{H}_\text{cc} \boldsymbol{S}_\text{cv}) (\boldsymbol{L}_{\text{vv}}^*)^{-1} , \\
\boldsymbol{\tilde{H}}_\text{vv} & = \boldsymbol{L}_{\text{vv}}^{-1} (\boldsymbol{H}_\text{vv} + \boldsymbol{S}_\text{vc} \boldsymbol{H}_\text{cc} \boldsymbol{S}_\text{cv} - \boldsymbol{H}_\text{vc} \boldsymbol{S}_\text{cv} - \boldsymbol{S}_\text{vc} \boldsymbol{H}_\text{cv}) (\boldsymbol{L}_{\text{vv}}^*)^{-1} . \label{eq:vv_ham}
\end{align}
\end{subequations}

\noindent Here $\boldsymbol{\tilde{H}}_\text{cc}$, $\boldsymbol{\tilde{H}}_\text{vc}$, $\boldsymbol{\tilde{H}}_\text{cv}$, and $\boldsymbol{\tilde{H}}_\text{vv}$ denote the cc, vc, cv, and vv blocks of $\boldsymbol{\tilde{H}}$, respectively. $\boldsymbol{\tilde{H}}_\text{vv}$ is a reduced Hamiltonian defined for the valence orbitals. The space spanned by the core orbitals is effectively projected out. $\boldsymbol{\tilde{H}}_\text{vc}$ and $\boldsymbol{\tilde{H}}_\text{cv}$ do not enter the variational optimization of the valence orbitals. Therefore, equation~\ref{eq:evp_matrix} can be rewritten as two separate eigenproblems,
\begin{subequations}
\label{eq:split_gevp}
\begin{align}
\boldsymbol{H}_\text{cc} \boldsymbol{C}_\text{cc} & = \boldsymbol{S}_\text{cc} \boldsymbol{C}_\text{cc} \boldsymbol{\Sigma}_\text{cc} , \label{eq:split_cc_gevp} \\
\boldsymbol{\tilde{H}}_\text{vv} \boldsymbol{\tilde{C}}_\text{vv} & = \boldsymbol{\tilde{S}}_\text{vv} \boldsymbol{\tilde{C}}_\text{vv} \boldsymbol{\Sigma}_\text{vv} , \label{eq:split_vv_gevp}
\end{align}
\end{subequations}

\noindent where
\begin{equation}
\label{eq:vv_ovlp}
\boldsymbol{\tilde{S}}_\text{vv} = \boldsymbol{S}_\text{vv} - \boldsymbol{S}_\text{vc} \boldsymbol{S}_\text{cv} .
\end{equation}

\noindent Considering (i) and (ii), the solution of equation~\ref{eq:split_cc_gevp} is simply
\begin{subequations}
\label{eq:cc_eval_evec}
\begin{align}
\boldsymbol{\Sigma}_{\text{cc}}^{(i)} & = \boldsymbol{H}_{\text{cc}}^{(i,i)} , \label{eq:cc_eval} \\
\boldsymbol{C}_{\text{cc}}^{(i)} & = \boldsymbol{\hat{u}}_i . \label{eq:cc_evec}
\end{align}
\end{subequations}

\noindent $\boldsymbol{\Sigma}_{\text{cc}}^{(i)}$ and $\boldsymbol{\tilde{C}}_{\text{cc}}^{(i)}$ denote the eigenvalue and eigenvector, respectively, of the $i^{\text{th}}$ core Kohn-Sham orbital. $\boldsymbol{H}_{\text{cc}}^{(i,i)}$ is the $i^{\text{th}}$ diagonal element of $\boldsymbol{H}_\text{cc}$. $\boldsymbol{\hat{u}}_i$ is a unit vector, meaning that this Kohn-Sham orbital exclusively consists of one atomic orbital basis function, as if the atomic orbital was frozen.

After equation~\ref{eq:split_vv_gevp} is solved, the full solution of equation~\ref{eq:gevp} can be constructed as follows:
\begin{subequations}
\label{eq:eval_evec}
\begin{align}
\boldsymbol{\Sigma} & =
\begin{bmatrix}
\boldsymbol{\Sigma}_\text{cc} & \boldsymbol{0} \\
\boldsymbol{0} & \boldsymbol{\Sigma}_\text{vv}
\end{bmatrix} , \\
\boldsymbol{C} & =
\begin{bmatrix}
\boldsymbol{C}_\text{cc} & -\boldsymbol{S}_\text{cv} \boldsymbol{\tilde{C}}_\text{vv} \\
\boldsymbol{0} & \boldsymbol{\tilde{C}}_\text{vv}
\end{bmatrix} . \label{eq:evec}
\end{align}
\end{subequations}

The FC99 method~\cite{fplo_koepernik_1999} further assumes the following relationship between the vc, cv and cc blocks of $\boldsymbol{H}$ and $\boldsymbol{S}$,
\begin{subequations}
\label{eq:vc_cv_blocks}
\begin{align}
\boldsymbol{H}_\text{vc} & = \boldsymbol{S}_\text{vc} \boldsymbol{H}_\text{cc} , \\
\boldsymbol{H}_\text{cv} & = \boldsymbol{H}_\text{cc} \boldsymbol{S}_\text{cv} .
\end{align}
\end{subequations}

\noindent The valence Hamiltonian is simplified to
\begin{equation}
\label{eq:vv_ham_reduced}
\boldsymbol{\tilde{H}}_\text{vv} = \boldsymbol{L}_{\text{vv}}^{-1} (\boldsymbol{H}_\text{vv} - \boldsymbol{S}_\text{vc} \boldsymbol{H}_\text{cc} \boldsymbol{S}_\text{cv}) (\boldsymbol{L}_{\text{vv}}^*)^{-1} .
\end{equation}

\noindent The impact of this approximation on the accuracy of the valence Kohn-Sham orbitals is investigated in section~\ref{sec:plus_v}.

Equations~\ref{eq:cholesky}--\ref{eq:eval_evec} do not hold if the frozen core atomic orbitals are not strictly orthonormalized. For general KS-DFT codes with non-orthogonal basis functions, technical implementation choices in different codes determine whether orthonormality is achieved exactly or only to a sufficiently good approximation, e.g., due to finite numerical integration grids. We here consider the FHI-aims software package~\cite{fhiaims_blum_2009}, a full-potential, all-electron implementation of electronic structure theory. FHI-aims uses atomic orbitals solved from a set of spherically symmetric free-atom Kohn-Sham equations as its basis functions. Thanks to the spherical symmetry, it suffices to solve the free-atom Kohn-Sham equations on a numerically highly precise one-dimensional (1D) logarithmic grid ($r_{i+1} = \alpha r_i$, where $r_{i+1}$ and $r_i$ are the $(i+1)^\text{th}$ and the $i^\text{th}$ points of the grid and $\alpha$ is a constant slightly larger than one, $\alpha=1.0123$ by default in FHI-aims) by using an atomic Kohn-Sham solver~\cite{atomic_fuchs_1999}. The atomic orbitals are orthonormalized on the 1D grid, then interpolated onto a three-dimensional (3D) overlapping atom-centered integration grid. Specifically, this grid (typical of exchange-correlation integrals in quantum chemistry\cite{quadrature_becke_1988}, but used for all numerical integrations in FHI-aims) consists of spherical shells of grid points, centered around each atom, but covering a much sparser set of radii than the 1D logarithmic grid. On each sphere, integration points are distributed to ensure a sufficiently accurate quadrature on a sphere, following prescriptions by Lebedev and by Delley~\cite{quadrature_lebedev_1975,quadrature_lebedev_1976,quadrature_lebedev_1999,quadrature_delley_1996}. A visualization of the 3D integration grid is available in, e.g., reference~\cite{gpu_huhn_2020}. Numerical errors are unavoidably introduced from the interpolation and the finite number of points in the 3D integration grid, though these errors can be systematically reduced by using a denser grid. This means that conditions (i) and (ii) mentioned at the beginning of this section will not be fulfilled exactly on the 3D grid.

\begin{figure}
\includegraphics[width=0.08\textwidth]{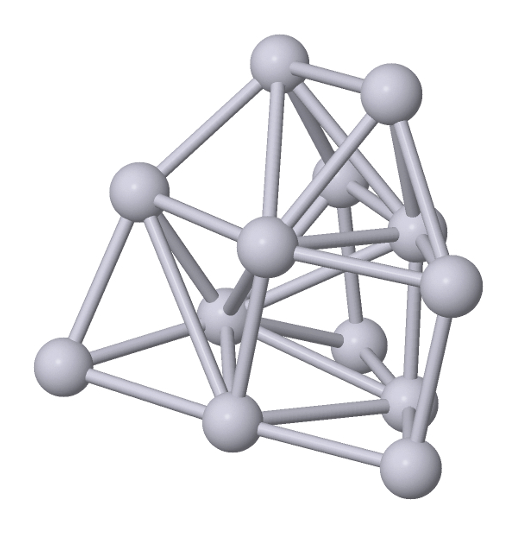}
\caption{Atomic structure of the Pt$_{13}$ cluster used for benchmarks in this work.}
\label{fig:pt13}
\end{figure}

We assess the impact of different 3D integration grids (particularly related to the density of radial grid shells in the 3D grids near the nucleus) on numerically integrated energies $\braket{\phi_i|\hat{h}^{\text{KS}}_{\text{atom}}|\phi_i}$ of core level basis functions of Pt in table~\ref{tab:integ_error}. The test computations use the FHI-aims code, the atomic zero-order regular approximation (ZORA) for relativistic effects~\cite{fhiaims_blum_2009}, and the PBE exchange-correlation functional~\cite{pbe_perdew_1996}. Two predefined numerical settings of FHI-aims, namely ``light'' and ``tight'', are tested. The ``light'' settings are intended for fast computations with acceptable precision for tasks such as preliminary atomic structure optimization or conformational sampling, while the ``tight'' settings provide high precision at a higher computational cost~\cite{fhiaims_blum_2009}. The ``light'' and ''tight'' settings mainly differ in the density of grid points, the size of the basis set, and the extent of the basis functions in real space. Table~\ref{tab:integ_error} shows that the 3D grid reduces the accuracy especially for the lowest-lying core level energies of Pt atoms (1s and 2s) and for ``light'' settings. Errors in these terms cancel in energy differences of different Pt containing structures and do not significantly affect the valence properties in all-electron calculations. Table~\ref{tab:integ_error} also includes the integration error $e_\text{s}$ of core basis function overlap matrix elements, obtained for a Pt$_{13}$ cluster, the structure of which is shown in figure~\ref{fig:pt13}. The integration errors for overlap matrix elements are seemingly much smaller than those of the Hamiltonian elements. However, their impact on calculated energies can still be significant since the core level energies themselves are of large magnitude with a negative sign (an example is given in section~\ref{sec:plus_c}).
\begin{table}
\caption{Integration error of quantities related to the core radial functions of Pt, comparing the 3D radial grid relative to the numerically highly precise 1D logarithmic grid. $e_\text{h}$: The mean absolute error (MAE) of $\braket{\phi_i|\hat{h}^{\text{KS}}_{\text{atom}}|\phi_i}$ computed on the 3D grid, compared to the same integral computed on the 1D grid. $\hat{h}^{\text{KS}}_{\text{atom}}$ is the free-atom Kohn-Sham Hamiltonian. $\phi_i$ is an atomic orbital. $e_\text{s}$: The MAE of $\braket{\phi_i|\phi_i}$ computed on the 3D grid compared to the theoretical value (one). For the overlap matrix elements, the test system is the Pt$_{13}$ cluster in figure~\ref{fig:pt13}. The errors with the ``light'' and ``tight'' numerical settings of FHI-aims are shown in atomic units.}
\begin{tabular}{c c c c c}
\hline
\hline
\multirow{2}{*}{state} & \multicolumn{2}{c}{$e_\text{h}$} & \multicolumn{2}{c}{$e_\text{s}$} \\
\cline{2-5}
& ``light'' & ``tight'' & ``light'' & ``tight'' \\
\hline
1s & $1.26 \times 10^{-2}$ & $7.64 \times 10^{-4}$ & $4.01 \times 10^{-6\z}$ & $2.42 \times 10^{-7\z}$ \\
2s & $2.44 \times 10^{-4}$ & $1.53 \times 10^{-5}$ & $4.65 \times 10^{-7\z}$ & $2.85 \times 10^{-8\z}$ \\
2p & $5.55 \times 10^{-7}$ & $3.23 \times 10^{-7}$ & $3.24 \times 10^{-10}$  & $1.38 \times 10^{-10}$  \\
3s & $1.25 \times 10^{-5}$ & $1.11 \times 10^{-6}$ & $1.03 \times 10^{-7\z}$ & $7.12 \times 10^{-8\z}$ \\
3p & $2.44 \times 10^{-7}$ & $2.27 \times 10^{-7}$ & $7.93 \times 10^{-10}$  & $5.74 \times 10^{-10}$  \\
3d & $9.81 \times 10^{-8}$ & $8.20 \times 10^{-8}$ & $1.94 \times 10^{-10}$  & $1.90 \times 10^{-10}$  \\
\hline
\hline
\end{tabular}
\label{tab:integ_error}
\end{table}

\begin{figure*}
\includegraphics[width=0.9\textwidth]{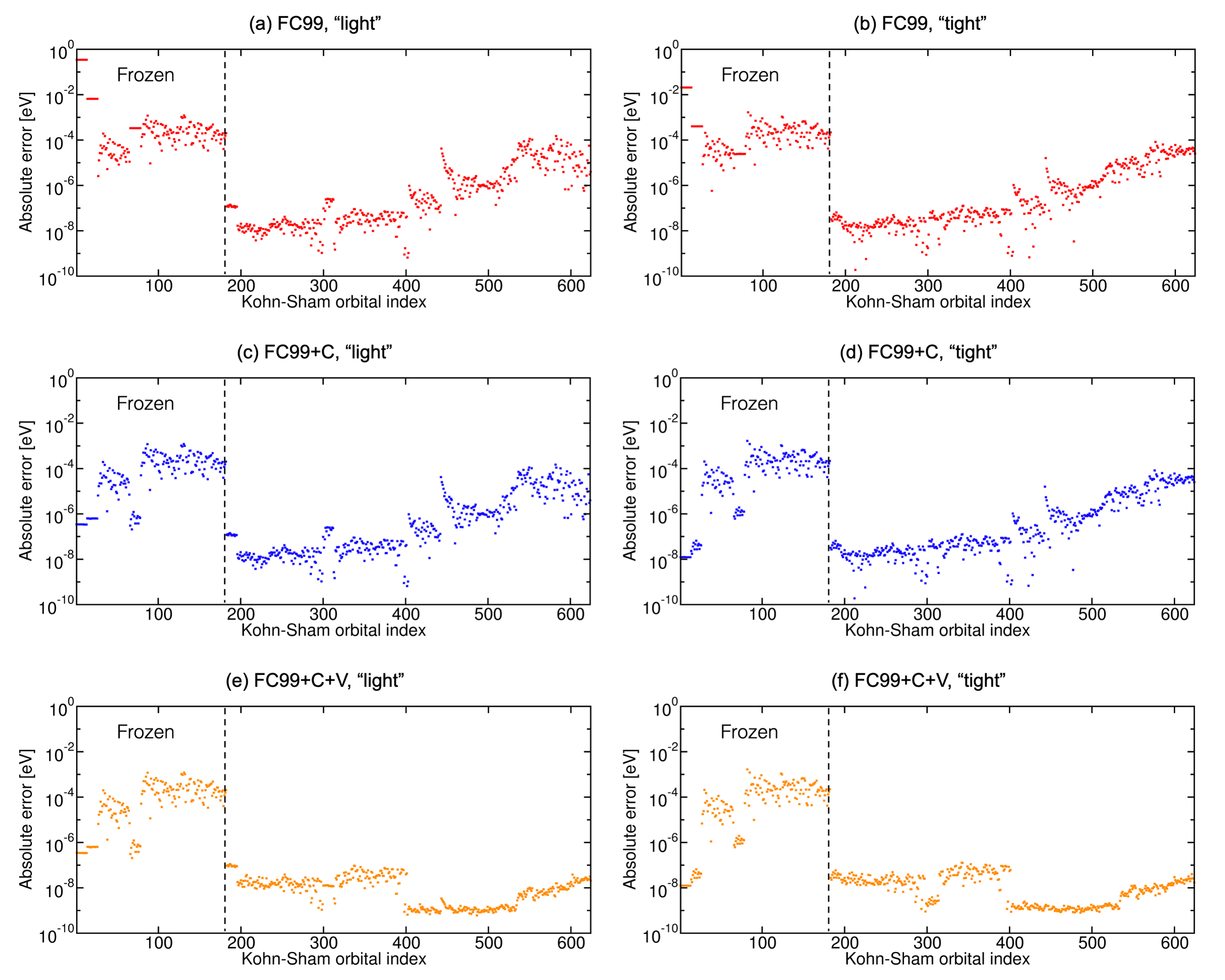}
\caption{Absolute error (relative to all-electron) of the eigenvalues of the Kohn-Sham orbitals computed with frozen core approximations. The test system is the Pt$_{13}$ cluster in figure~\ref{fig:pt13}, with the 1s, 2s, 2p, 3s, 3p, and 3d orbitals of Pt frozen. The dashed vertical line in each figure separates frozen and unfrozen orbitals. (a) and (b): The frozen core approximation proposed by Koepernik and Eschrig~\cite{fplo_koepernik_1999} (FC99). (c) and (d): The core-corrected frozen core approximation in this work (FC99+C). (e) and (f): The core-and-valence-corrected frozen core approximation in this work (FC99+C+V). (a), (c), (e) use the ``light'' numerical settings of FHI-aims. (b), (d), (f) use the ``tight'' settings.}
\label{fig:eval}
\end{figure*}

Figure~\ref{fig:eval} presents the absolute error of the eigenvalues of the Kohn-Sham orbitals of the Pt$_{13}$ cluster in figure~\ref{fig:pt13} when freezing the 1s, 2s, 2p, 3s, 3p, and 3d orbitals of the Pt atoms, compared to all-electron reference values. As shown in figure~\ref{fig:eval} (a) and (b), with the ``light'' (``tight'') numerical settings, the FC99 method alters the eigenvalues by as much as 0.3 (0.02) eV. Returning to the integration errors $e_\text{h}$ in table~\ref{tab:integ_error}, for core orbitals, $\braket{\phi_i|\hat{h}^{\text{KS}}_{\text{atom}}|\phi_i} \approx \braket{\phi_i|\hat{h}^\text{KS}|\phi_i} = \boldsymbol{H}_{\text{cc}}^{(i,i)}$. Thus, the matrix elements of the core Hamiltonian matrix of the Pt$_{13}$ cluster computed on the 3D grid would not be highly accurate. The eigenvalues computed via equation~\ref{eq:cc_eval} would not be accurate either, as seen in figure~\ref{fig:eval} (a) and (b).

\begin{figure}
\includegraphics[width=0.45\textwidth]{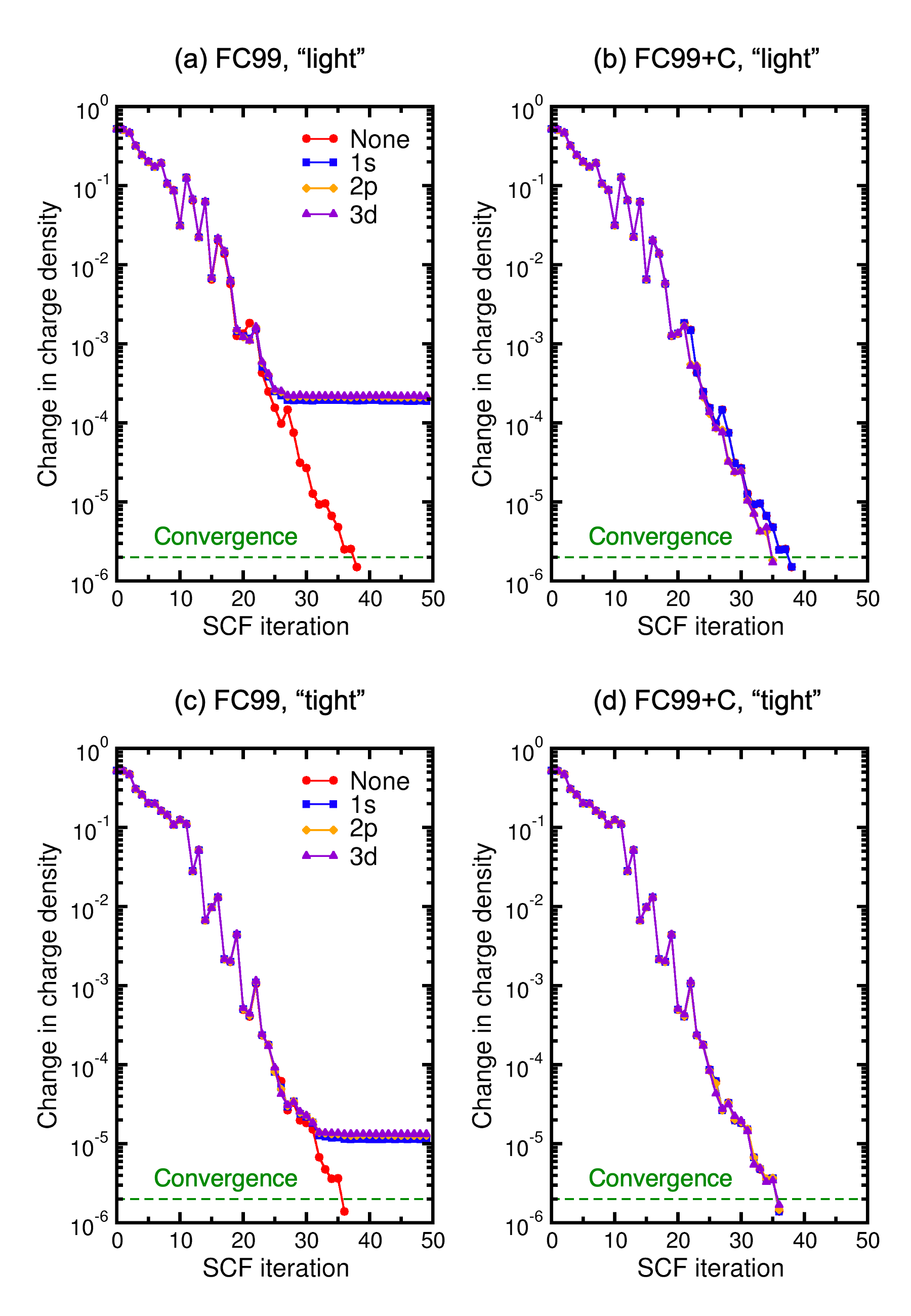}
\caption{Change in the norm of the charge density as a function of the SCF iteration number for the Pt$_{13}$ cluster in figure~\ref{fig:pt13}. The convergence criterion is $2 \times$ 10$^{-6}$. (a) and (c): FC99. (b) and (d): FC99+C. (a) and (b): The ``light'' numerical settings of FHI-aims. (c) and (d): The ``tight'' numerical settings of FHI-aims. Red circles: all-electron. Blue squares: 1s orbitals frozen. Orange diamonds: 1s, 2s, and 2p orbitals frozen. Violet triangles: 1s, 2s, 2p, 3s, 3p, and 3d orbitals frozen.}
\label{fig:scf}
\end{figure}

Using the FC99 method with the errors in table~\ref{tab:integ_error} also introduces errors to the eigenvectors (i.e. Kohn-Sham orbitals) computed via equation~\ref{eq:cc_evec}. As mentioned in section~\ref{sec:intro}, the solution of the Kohn-Sham equations is a non-linear optimization problem, and it is typically done in an SCF procedure. In most, if not all KS-DFT codes, the SCF procedure is accelerated by ``mixing'' techniques such as Pulay's method~\cite{pulay_pulay_1980}. The convergence can usually be achieved within tens of iterations. However, errors in the Kohn-Sham orbitals enter the electron density via equation~\ref{eq:density_from_orbital}, resulting in a small numerical uncertainty that prevents the non-linear optimization from converging. As an example, figure~\ref{fig:scf} (a) and (c) show the change in the norm of the electron density as the SCF cycle proceeds for the Pt$_{13}$ cluster in figure~\ref{fig:pt13}, comparing all-electron and frozen core (FC99) calculations with the FHI-aims code. Pulay's mixer is used in both cases. The all-electron calculations converge to the prescribed threshold, $2 \times 10^{-6}$, in fewer than 40 iterations. In contrast, the FC99 calculations stall at around 10$^{-4}$ (10$^{-5}$) with the ``light'' (``tight'') settings even if only the 1s orbitals are frozen. While the deviation of the almost converged SCF cycle from the fully converged result is physically negligible, it is striking that even small normalization errors can affect the convergence of the non-linear fixed-point iteration used to converge the SCF cycle. In contrast, the correct normalization (which is achieved through the generalized eigensolver in the all-electron case) poses no such issues.

\subsection{Core Correction}
\label{sec:plus_c}
\begin{figure}
\includegraphics[width=0.5\textwidth]{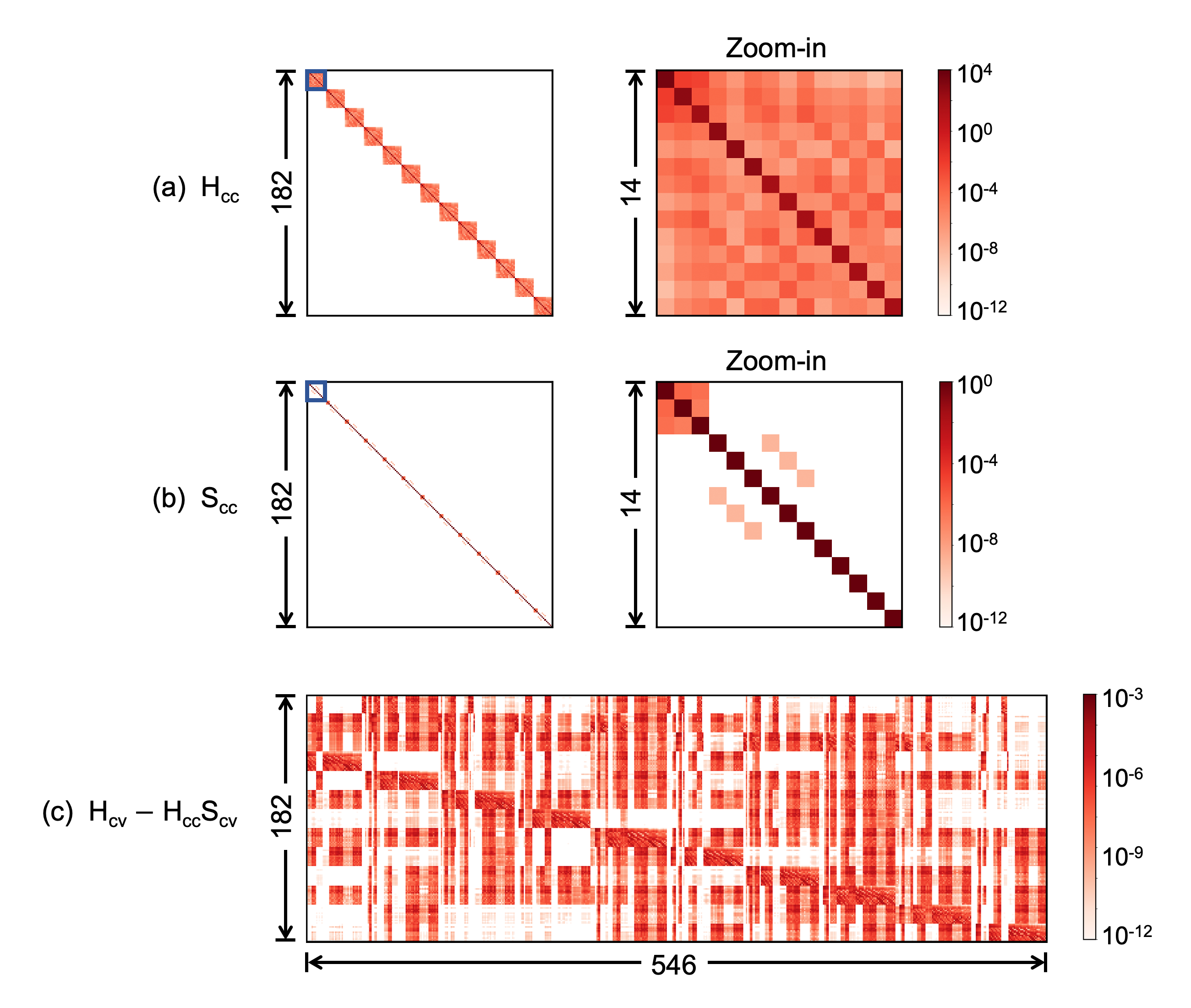}
\caption{Visualization of the matrix elements (absolute values) of (a) $\boldsymbol{H}_\text{cc}$, (b) $\boldsymbol{S}_\text{cc}$, and (c) $\boldsymbol{H}_\text{cv} - \boldsymbol{H}_\text{cc} \boldsymbol{S}_\text{cv}$, obtained from the first SCF iteration of KS-DFT calculation of the Pt$_{13}$ cluster in figure~\ref{fig:pt13}, using the PBE functional, the FHI-aims code, and the ``light'' numerical settings of FHI-aims. The 1s, 2s, 2p, 3s, 3p, and 3d orbitals of the Pt atoms are frozen, yielding 14 frozen orbitals per Pt atom and 182 frozen orbitals in total. The number of unfrozen orbitals is 546. The right column of (a) and (b) shows the zoom-in of one atomic block of $\boldsymbol{H}_\text{cc}$ and $\boldsymbol{S}_\text{cc}$, corresponding to the boxes in the left column.}
\label{fig:mat}
\end{figure}

Figure~\ref{fig:mat} (a) and (b) visualize the $\boldsymbol{H}_\text{cc}$ and $\boldsymbol{S}_\text{cc}$ matrices of the Pt$_{13}$ cluster, obtained with the ``light'' settings of FHI-aims, in the first SCF iteration, and with the 1s, 2s, 2p, 3s, 3p, and 3d orbitals frozen. Neither $\boldsymbol{H}_\text{cc}$ nor $\boldsymbol{S}_\text{cc}$ is diagonal. Again, this is a consequence of the finite precision of the 3D integration grid and the fact that the basis functions are not orthonormalized on the 3D grid. Additional tests (not shown here for simplicity) suggest that freezing more core orbitals yields more off-diagonal elements in $\boldsymbol{H}_\text{cc}$ and $\boldsymbol{S}_\text{cc}$, and using a denser integration grid, e.g. the ``tight'' settings, lowers the magnitude of the off-diagonal elements.

Although not strictly diagonal, $\boldsymbol{H}_\text{cc}$ and $\boldsymbol{S}_\text{cc}$ are both diagonally-dominant. Their off-diagonal elements are more than four orders of magnitude smaller than the diagonal elements, as shown in the zoom-in of one atomic block of $\boldsymbol{H}_\text{cc}$ and $\boldsymbol{S}_\text{cc}$. Therefore, instead of treating $\boldsymbol{S}_\text{cc}$ as an identity matrix, we treat it as a general diagonal matrix with diagonal elements close to but not necessarily equal to one. The square root of $\boldsymbol{S}_\text{cc}$ is then computed by taking the square root of its diagonal elements,
\begin{equation}
\label{eq:cc_ovlp_sqrt}
\boldsymbol{S}_\text{cc}^{1/2} =
\begin{bmatrix}
\sqrt{\boldsymbol{S}_{\text{cc}}^{(1,1)}} & & & \\
& \sqrt{\boldsymbol{S}_{\text{cc}}^{(2,2)}} & & \\
& & \cdots & \\
& & & \sqrt{\boldsymbol{S}_{\text{cc}}^{(N_\text{c},N_\text{c}})}
\end{bmatrix} ,
\end{equation}

\noindent where $\boldsymbol{S}_{\text{cc}}^{(i,i)}$ is the $i^\text{th}$ diagonal element of $\boldsymbol{S}_\text{cc}$. Using $\boldsymbol{S}_\text{cc}^{1/2}$, the core block of equation~\ref{eq:gevp} is transformed to a standard eigenproblem
\begin{equation}
\label{eq:cc_evp}
\Big (\boldsymbol{S}_\text{cc}^{-1/2} \boldsymbol{H}_\text{cc} \boldsymbol{S}_\text{cc}^{-1/2} \Big ) \Big (\boldsymbol{S}_\text{cc}^{1/2} \boldsymbol{C}_\text{cc} \Big ) = \Big (\boldsymbol{S}_\text{cc}^{1/2} \boldsymbol{C}_\text{cc} \Big ) \boldsymbol{\Sigma}_\text{cc} .
\end{equation}

\noindent Accordingly, equation~\ref{eq:cc_eval_evec} becomes
\begin{subequations}
\label{eq:cc_eval_evec_plus_c}
\begin{align}
\boldsymbol{\Sigma}_{\text{cc}}^{(i)} & = \frac{\boldsymbol{H}_{\text{cc}}^{(i,i)}}{\boldsymbol{S}_{\text{cc}}^{(i,i)}} , \label{eq:cc_eval_plus_c} \\
\boldsymbol{C}_{\text{cc}}^{(i)} & = \frac{\boldsymbol{\hat{u}}_i}{\sqrt{\boldsymbol{S}_{\text{cc}}^{(i,i)}}} . \label{eq:cc_evec_plus_c}
\end{align}
\end{subequations}

\noindent Equation~\ref{eq:cc_evec_plus_c} results from the orthonormality condition, $\boldsymbol{C}_{\text{cc}}^* \boldsymbol{S}_{\text{cc}} \boldsymbol{C}_{\text{cc}} = \boldsymbol{I}$. A formula similar to equation~\ref{eq:cc_eval_evec_plus_c} (hereafter referred to as the core correction, ``+C'') was derived by L\"{o}wdin in a different context~\cite{overlap_lowdin_1950}. In the early days of the linear combination of atomic orbitals (LCAO) theory, it was common to adopt $\boldsymbol{S} = \boldsymbol{I}$ in equation~\ref{eq:gevp} due to the limit in computational resources. L\"{o}wdin noted that incorporating the overlap matrix into equation~\ref{eq:gevp} leads to a correction similar to equation~\ref{eq:cc_eval_evec_plus_c}.

Let $\boldsymbol{S}_{\text{cc}}^{(i,i)} = 1 - e_\text{s}$, then by Taylor expansion, $\boldsymbol{\Sigma}_{\text{cc}}^{(i)} = \boldsymbol{H}_{\text{cc}}^{(i,i)} + e_\text{s} \boldsymbol{H}_{\text{cc}}^{(i,i)} + \cdots$. The magnitude of $e_\text{s}$ is listed in table~\ref{tab:integ_error}. Omitting $e_\text{s}$ (equation~\ref{eq:cc_eval_evec}) introduces a first order error of $e_\text{s} \boldsymbol{H}_{\text{cc}}^{(i,i)}$, which, depending on the magnitude of $\boldsymbol{H}_{\text{cc}}^{(i,i)}$, can become very large. This error is eliminated by using equation~\ref{eq:cc_eval_plus_c} instead of equation~\ref{eq:cc_eval}. As shown in figure~\ref{fig:eval} (c) and (d), the +C correction significantly improves the accuracy of the frozen s orbitals. It has very little impact on the frozen p and d orbitals, since $e_\text{s}$ is relatively small for these orbitals (see table~\ref{tab:integ_error}).

The accuracy of the eigenvectors, i.e., Kohn-Sham orbitals, is also improved by equation~\ref{eq:cc_evec_plus_c}. The computation of the electron density, equation~\ref{eq:density_from_orbital}, requires properly orthonormalized orbitals. Errors in the electron density, originating from badly orthonormalized orbitals, can prevent the SCF cycle from converging, as exemplified by figure~\ref{fig:scf} (a) and (c). An explicit orthonormalization can be done, but it is computationally expensive with an $\mathcal{O}(N^3)$ cost. We define
\begin{subequations}
\label{eq:orth_norm_error}
\begin{align}
e_\text{orth} & = \frac{2}{N_\text{basis} (N_\text{basis}-1)} \sum_{i=1}^{N_\text{basis}} \sum_{j=i+1}^{N_\text{basis}} \vert \boldsymbol{C}_i^* \boldsymbol{S} \boldsymbol{C}_j \vert, \\
e_\text{norm} & = \frac{1}{N_\text{basis}} \sum_{i=1}^{{N_\text{basis}}} \vert \boldsymbol{C}_i^* \boldsymbol{S} \boldsymbol{C}_i - 1 \vert .
\end{align}
\end{subequations}

\noindent to measure the orthonormality errors of the eigenvectors. Here $\boldsymbol{S}$ is the overlap matrix out of equation~\ref{eq:ham_integ}. $\boldsymbol{C}_i$ and $\boldsymbol{C}_j$ denote the $i^{\text{th}}$ and $j^{\text{th}}$ eigenvectors, obtained by directly solving equation~\ref{eq:gevp} in the all-electron case, by using equations~\ref{eq:cc_evec} and \ref{eq:evec} in the FC99 case, and by using equations~\ref{eq:evec} and \ref{eq:cc_evec_plus_c} in the FC99+C case. Table~\ref{tab:orth_error} lists $e_\text{orth}$ and $e_\text{norm}$ in the computation of the Pt$_{13}$ cluster. When not freezing any orbital, both $e_\text{orth}$ and $e_\text{norm}$ are below $10^{-14}$. When using FC99 or FC99+C, $e_\text{orth}$ is still reasonably small. However, when using FC99, freezing only the 1s orbitals of Pt drastically increases $e_\text{norm}$ to $1.03 \times 10^{-7}$ ($6.20 \times 10^{-9}$) with the ``light'' (``tight'') settings. FC99+C brings $e_\text{norm}$ down to $1.90 \times 10^{-14}$ ($2.22 \times 10^{-15}$) with the ``light'' (``tight'') settings, which turns out to resolve the SCF convergence issue. In figure~\ref{fig:scf} (b) and (d), the convergence of FC99+C is as good as that of the all-electron calculations.
\begin{table}
\caption{Orthogonality errors $e_\text{orth}$ and $e_\text{norm}$, defined in equation~\ref{eq:orth_norm_error}, in all-electron (AE), FC99, and FC99+C calculations. The test system is the Pt$_{13}$ cluster in figure~\ref{fig:pt13}. The 1s orbitals of Pt are frozen. The errors with the ``light'' and ``tight'' numerical settings of FHI-aims are shown.}
\begin{tabular}{l c c c c}
\hline
\hline
\multirow{2}{*}{method} & \multicolumn{2}{c}{$e_\text{orth}$} & \multicolumn{2}{c}{$e_\text{norm}$} \\
\cline{2-5}
& ``light'' & ``tight'' & ``light'' & ``tight'' \\
\hline
AE     & $5.72 \times 10^{-17}$ & $5.21 \times 10^{-17}$ & $9.45 \times 10^{-16}$  & $1.25 \times 10^{-15}$ \\
FC99   & $8.31 \times 10^{-13}$ & $4.84 \times 10^{-14}$ & $1.03 \times 10^{-7\z}$ & $6.20 \times 10^{-9\z}$ \\
FC99+C & $8.31 \times 10^{-13}$ & $4.84 \times 10^{-14}$ & $1.90 \times 10^{-14}$  & $2.22 \times 10^{-15}$ \\
\hline
\hline
\end{tabular}
\label{tab:orth_error}
\end{table}

\subsection{Valence Correction}
\label{sec:plus_v}
The FC99 method assumes that $\boldsymbol{H}_\text{cv} - \boldsymbol{H}_\text{cc} \boldsymbol{S}_\text{cv} = \boldsymbol{0}$, so that equation~\ref{eq:vv_ham} is reduced to equation~\ref{eq:vv_ham_reduced}. Figure~\ref{fig:mat} (c) visualizes $\boldsymbol{H}_\text{cv} - \boldsymbol{H}_\text{cc} \boldsymbol{S}_\text{cv}$ of the Pt$_{13}$ cluster obtained with the ``light'' settings of FHI-aims and with the 1s, 2s, 2p, 3s, 3p, and 3d orbitals frozen. $\boldsymbol{H}_\text{cv} - \boldsymbol{H}_\text{cc} \boldsymbol{S}_\text{cv}$ contains a number of non-zero values (absolute value $< 10^{-3}$), suggesting that replacing equation~\ref{eq:vv_ham} with equation~\ref{eq:vv_ham_reduced} would cause non-negligible errors. We therefore undo the approximation in equation~\ref{eq:vc_cv_blocks}, i.e., we use the more accurate equation~\ref{eq:vv_ham}. $\boldsymbol{S}_\text{vc} \boldsymbol{H}_\text{cc} \boldsymbol{S}_\text{cv} - \boldsymbol{H}_\text{vc} \boldsymbol{S}_\text{cv} - \boldsymbol{S}_\text{vc} \boldsymbol{H}_\text{cv}$ in equation~\ref{eq:vv_ham} can be computed as $\boldsymbol{A} + \boldsymbol{A}^*$, $\boldsymbol{A} = (0.5 \boldsymbol{S}_\text{vc} \boldsymbol{H}_\text{cc} - \boldsymbol{H}_\text{vc}) \boldsymbol{S}_\text{cv}$, which, compared to equation~\ref{eq:vv_ham_reduced}, increases the computational cost only marginally (both require two matrix multiplications). As shown in figure~\ref{fig:eval} (e) and (f), this correction (hereafter referred to as the valence correction, ``+V'') indeed improves the accuracy of the eigenvalues of the valence orbitals.

We note that the FC99, FC99+C, and FC99+C+V methods can be used in essentially any all-electron electronic structure codes employing localized basis sets, provided that accurate atomic orbitals for the core electrons are available.

\section{Accuracy}
\label{sec:accuracy}
We assess the accuracy of the FC99+C+V frozen core approximation using a set of 103 materials, containing chemical species from lithium to polonium in the periodic table. This test set was defined in reference~\cite{soc_huhn_2017} and used there for a comprehensive benchmark of energy band structures at different levels of approximations to relativity in DFT. The 103 materials covers ten typical crystal structures: face-centered cubic (FCC), body-centered cubic (BCC), simple cubic (SC), hexagonal close-packed (HCP), graphite (GRA), diamond (DIA), cubic zincblende (ZB), wurtzite (WUR), rocksalt (RS), and cesium chloride (CSCL). They are classified into three subsets: compound semiconductors (37 materials), elemental materials (45 materials), and alkali halides (21 materials), as listed below.
\begin{itemize}
\item Compound semiconductors: C (DIA), MgO (RS), AlN (WUR), AlN (ZB), SiC (ZB), BP (ZB), AlP (ZB), MgS (RS), ZnO (WUR), ZnS (WUR), ZnS (ZB), GaN (WUR), GaN (ZB), GaP (ZB), BAs (ZB), AlAs (ZB), GaAs (ZB), MgSe (RS), ZnSe (ZB), CdS (WUR), CdS (ZB), CdSe (WUR), CdSe (ZB), InN (WUR), InP (ZB), InAs (ZB), AlSb (ZB), GaSb (ZB), InSb (ZB), ZnTe (ZB), CdTe (ZB), HgS (ZB), HgSe (ZB), HgTe (ZB), PbS (RS), PbSe (RS), PbTe (RS).
\item Elemental materials: Be (HCP), C (GRA), Ne (FCC), Mg (HCP), Al (FCC), Si (DIA), Ca (FCC), Sc (HCP), Ti (HCP), V (BCC), Cr (BCC), Mn (FCC), Fe (BCC), Co (HCP), Ni (FCC), Cu (FCC), Zn (HCP), Ge (DIA), Sr (FCC), Y (HCP), Zr (HCP), Nb (BCC), Mo (BCC), Tc (HCP), Ru (HCP), Rh (FCC), Pd (FCC), Ag (FCC), Cd (HCP), Sn (DIA), Xe (FCC), Ba (BCC), Lu (HCP), Hf (HCP), Ta (BCC), W (BCC), Re (HCP), Os (HCP), Ir (FCC), Pt (FCC), Au (FCC), Tl (HCP), Pb (FCC), Bi (BCC), Po (SC).
\item Alkali halides: LiF (RS), NaF (RS), LiCl (RS), NaCl (RS), KF (RS), KCl (RS), LiBr (RS), NaBr (RS), KBr (RS), RbF (RS), RbCl (RS), RbBr (RS), LiI (RS), NaI (RS), KI (RS), RbI (RS), CsF (RS), CsCl (CSCL), CsCl (RS), CsBr (CSCL), CsI (CSCL).
\end{itemize}

\noindent The Delta test~\cite{delta_lejaeghere_2016} has in principle validated the efficacy of the frozen core approximation by comparing all-electron and pseudopotential DFT codes. However, all kinds of theoretical and technical aspects can be very different between the tested all-electron and pseudopotential codes. Our goal here is to isolate the effect of the frozen core approximation, without introducing any other variables to the test.

\begin{figure*}
\includegraphics[width=0.9\textwidth]{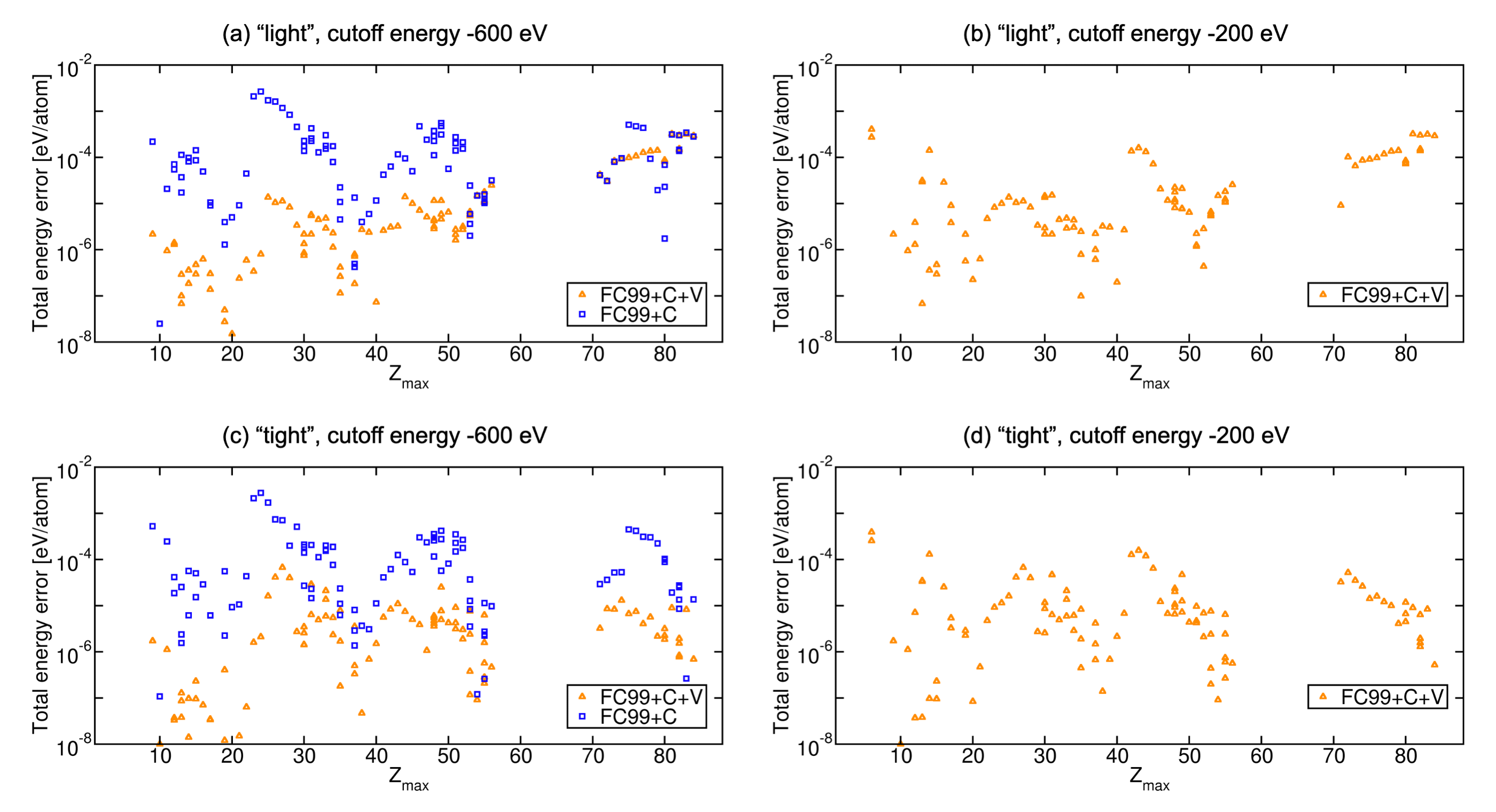}
\caption{Absolute error (relative to all-electron) of the total energies of the 103 materials computed with frozen core approximations. Blue squares: FC99+C. Orange triangles: FC99+C+V. (a) and (b) use the ``light'' numerical settings of FHI-aims. (c) and (d) use the ``tight'' settings. Frozen core cutoff energy $E_\text{cut}$ is $-600$ eV in (a) and (c), and $-200$ eV in (b) and (d).}
\label{fig:103}
\end{figure*}

Figure~\ref{fig:103} shows the total energy errors (relative to all-electron) of FC99+C and FC99+C+V for the 103 materials. The materials are sorted according to their maximum atomic number $Z_\text{max}$. The number of frozen orbitals is controlled by an energy cutoff parameter $E_\text{cut}$. Atomic orbitals with an eigenvalue smaller than $E_\text{cut}$ are frozen. While the absolute energy reference is arbitrary in periodic calculations, a detailed list of the frozen orbitals for all the materials is given in the supplementary material. For $E_\text{cut} = -600$ eV in figure~\ref{fig:103} (a) and (c), the total energies computed with the FC99+C method agree reasonably well with the all-electron values. The mean absolute error (MAE) over the 103 materials is $2.21 \times 10^{-4}$ ($1.77 \times 10^{-4}$) eV/atom with the ``light'' (``tight'') numerical settings of FHI-aims. The largest errors appear with materials containing transition metals. For the same $E_\text{cut}$, the errors of FC99+C+V are generally lower than those of FC99+C. The MAE is only $3.02 \times 10^{-5}$ (5.39 $\times 10^{-6}$) eV/atom with the ``light'' (``tight'') settings. For $E_\text{cut} = -200$ eV in figure~\ref{fig:103} (b) and (d), FC99+C+V is still highly accurate. The MAE is $4.49 \times 10^{-5}$ (2.16 $\times 10^{-5}$) eV/atom with the ``light'' (``tight'') settings, better than FC99+C with $E_\text{cut} = -600$ eV. In contrast, SCF convergence issues can occur in FC99+C with $E_\text{cut} = -200$ eV, and in FC99 regardless of the choice of $E_\text{cut}$. This is due to the use of the more approximate equation~\ref{eq:vv_ham_reduced}. The more accurate and more robust FC99+C+V method is therefore recommended for practical use. Even the maximum total energy error of the FC99+C+V approach with $E_\text{cut} = -200$ eV is still well below 10$^{-3}$ eV/atom. While the authors were aware that the frozen core approach is well justified, this very low error value is nevertheless remarkable, given the wide benchmark performed. The authors are not aware of a comparable benchmark of only the frozen core approximation, with all other factors being equal, for the total energy, i.e., the central quantity sought in DFT.

In tests with FC99+C+V and $E_\text{cut} = -100$ eV, the SCF cycle no longer converges well for around 30 cases out of the 103-material benchmark set. This is likely due to the expected breakdown of strict absence of overlap between frozen orbitals centered at different atoms in this case. As in the case of figure~\ref{fig:scf}, it is conceivable that most of these numerical SCF convergence issues are minor from a physical point of view and could perhaps be overcome by a more refined treatment of the SCF cycle. Nevertheless, the observations show that even the more robust method, FC99+C+V, begins to be affected by the small violations of orthonormality that arise for $E_\text{cut} = -100$ eV. Thus, a cutoff energy $E_\text{cut} = -200$ eV appears to be a safe choice in practical applications of FC99+C+V.

\section{Performance}
\label{sec:performance}
\begin{figure}
\includegraphics[width=0.15\textwidth]{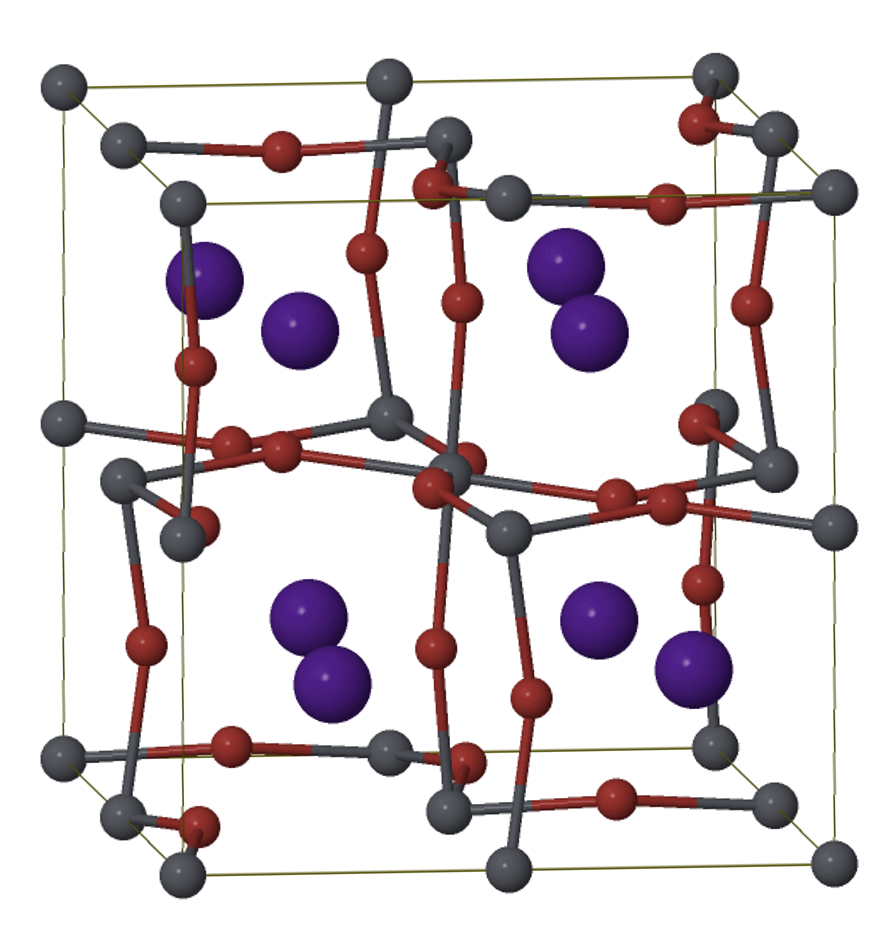}
\caption{Atomic structure of the CsPbBr$_3$ model. Violet: Cs. Gray: Pb. Red: Br. There are 8 Cs atoms, 8 Pb atoms, and 24 Br atoms in the unit cell. Lattice parameters a = 11.356 \AA, b = 11.848 \AA, c = 11.356 \AA, $\alpha$ = 90$^\circ$, $\beta$ = 85.003$^\circ$, $\gamma$ = 90$^\circ$.}
\label{fig:cspbbr3}
\end{figure}

In this section we examine the computational efficiency of the FC99+C+V frozen core approximation. We choose CsPbBr$_3$, a perovskite consisting of heavy elements with a high number of core orbitals, as the test system. CsPbBr$_3$ is made of caesium (atomic number Z = 59), lead (Z = 82), and bromine (Z = 35). Its atomic structure is shown in figure~\ref{fig:cspbbr3}. Unless stated otherwise, the computations in this section use a $4 \times 4 \times 4$ $\boldsymbol{k}$-point grid, the PBE exchange-correlation functional~\cite{pbe_perdew_1996}, the atomic ZORA treatment of relativity~\cite{fhiaims_blum_2009}, and the Cray XC40 supercomputer Cori (Haswell partition) at National Energy Research Scientific Computing Center (NERSC). Each node of Cori-Haswell is equipped with two 16-core Intel Haswell CPU processors. The valence eigenproblem, equation~\ref{eq:split_vv_gevp}, is solved by the ELPA eigensolver library~\cite{elpa_auckenthaler_2011,elpa_marek_2014}.

Table~\ref{tab:time} lists the total energy error of the FC99+C+V method relative to the all-electron results. The error generally lies below one meV/atom. Freezing the 1s orbitals of Cs and Br and the 1s, 2s, and 2p orbitals of Pb only changes the total energy by 43.8 (0.2) $\mu$eV/atom with the ``light'' (``tight'') settings of FHI-aims. Freezing orbitals up to 4p for Cs, 5s for Pb, and 3p for Br still only changes the total energy by 75.2 (16.3) $\mu$eV/atom with the ``light'' (``tight'') settings. Also shown in table~\ref{tab:time} is the time spent on the solution of the generalized eigenproblem, equation~\ref{eq:gevp}. Freezing core orbitals has a clear advantage in computational efficiency. For instance, freezing 448 orbitals, i.e., 30.4\% (25.6\%) of the basis functions in the ``light'' (``tight'') settings, leads to a speedup of 2.0x (1.8x). In practice, the accuracy and computational cost of the method can be balanced by tuning the number of frozen orbitals.
\begin{table}
\caption{Total energy error (relative to all-electron) and run time of the FC99+C+V method. The test system is the CsPbBr$_3$ model in figure~\ref{fig:cspbbr3}. $N_\text{core}$ is the number of frozen orbitals. $N_\text{basis} = 1,472$ with the ``light'' settings. $N_\text{basis} = 1,752$ with the ``tight'' settings. A $4 \times 4 \times 4$ $\boldsymbol{k}$-grid is used. $\Delta E_\text{tot} = \vert E_{\text{FC}} - E_\text{AE} \vert$, where $E_{\text{FC}}$ and $E_\text{AE}$ are the total energies computed with and without the FC99+C+V frozen core approximation. $t_\text{eigen}$ is the time to solve equation~\ref{eq:gevp} using one node of Cori-Haswell.}
\footnotesize
\begin{tabular}{c c c c c c c c c}
\hline
\hline
Cs & Pb & Br & & & \multicolumn{2}{c}{``light''} & \multicolumn{2}{c}{``tight''} \\
\cline{1-3} \cline{6-9}
\multicolumn{3}{c}{frozen orbitals} & $E_\text{cut}$ [eV] & $N_\text{core}$ & $\Delta E_\text{tot}$ & $t_\text{eigen}$ & $\Delta E_\text{tot}$ & $t_\text{eigen}$ \\
\multicolumn{3}{c}{(up to)} & & & [$\mu$eV/atom] & [s] & [$\mu$eV/atom] & [s] \\
\hline
   &    &    &          & \z\z0 & \z0.0 & 4.5 & \z0.0 & 7.1 \\
1s & 2p & 1s &   -13000 &  \z72 &  43.8 & 4.2 & \z0.2 & 6.8 \\
2p & 3d & 2p &  \z-1500 &   272 &  59.0 & 3.0 & \z0.2 & 5.2 \\
3d & 4p & 2p & \z\z-600 &   376 &  62.2 & 2.5 & \z0.2 & 4.4 \\
4s & 4d & 3s & \z\z-200 &   448 &  63.0 & 2.2 & \z0.7 & 3.9 \\
4p & 5s & 3p & \z\z-100 &   608 &  75.2 & 1.7 &  16.3 & 3.0 \\
\hline
\hline
\end{tabular}
\label{tab:time}
\end{table}

\begin{figure}
\includegraphics[width=0.35\textwidth]{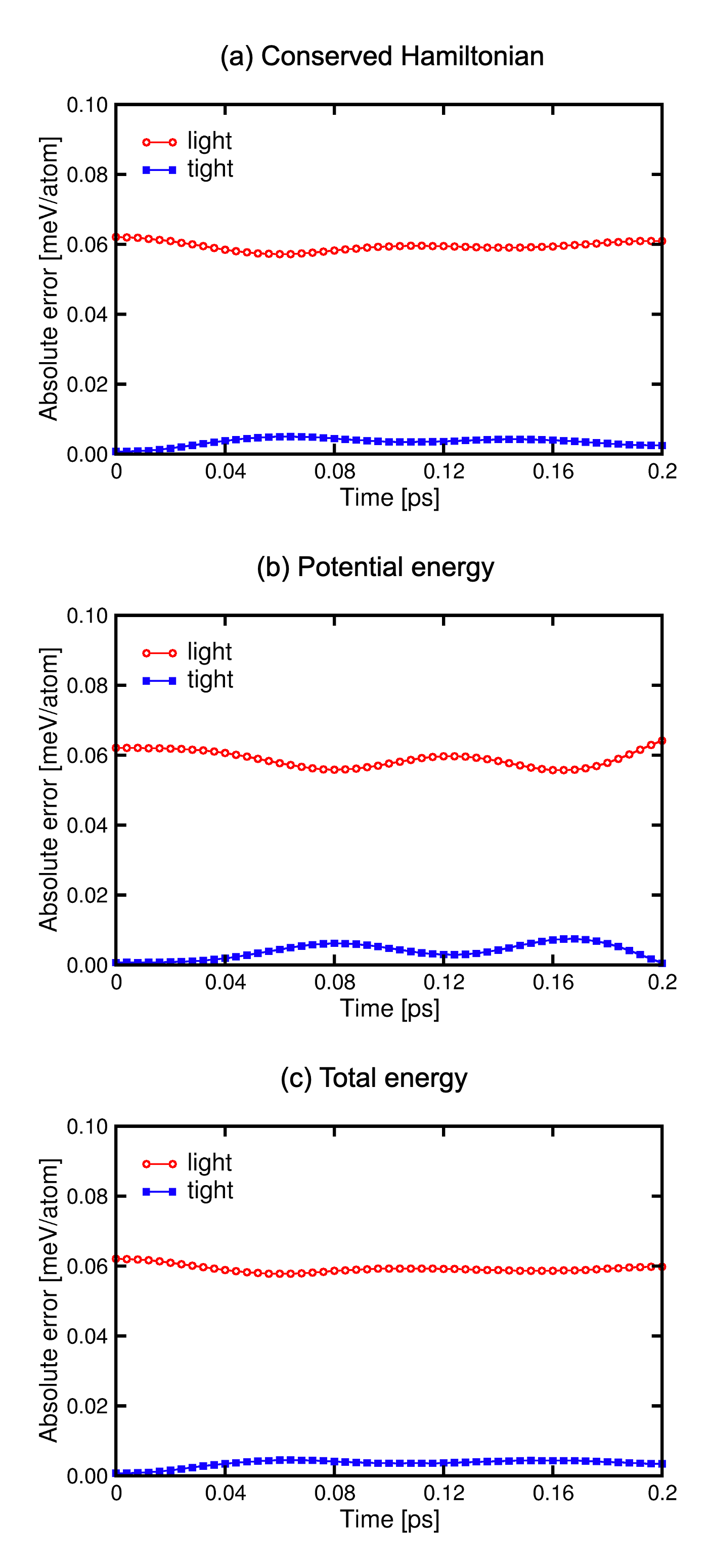}
\caption{Absolute error of FC99+C+V (relative to all-electron) as a function of the simulation time (total 0.2 ps, timestep 0.004 ps). The test system is the CsPbBr$_3$ model in figure~\ref{fig:cspbbr3}. (a) Conserved pseudo-Hamiltonian. (b) Potential energy. (c) Total energy. Red circles: ``light'' settings. Blue squares: ``tight'' settings. Orbitals up to the 4s, 4d, and 3s states are frozen for the Cs, Pb, and Br atoms, respectively.}
\label{fig:md}
\end{figure}

To check the accuracy of the forces computed with FC99+C+V, we perform molecular dynamics simulations in the canonical ensemble using the Bussi-Donadio-Parrinello (BDP) thermostat~\cite{md_bussi_2007}. A timestep of 0.004 ps is used. The BDP method has a conserved quantity, the pseudo Hamiltonian, throughout the simulation. The error of FC99+C+V relative to all-electron, as a function of the simulation time (first 0.2 ps), is plotted in figure~\ref{fig:md}. In the FC99+C+V calculation, the orbitals up to the 4s, 4d, and 3s states are frozen for the Cs, Pb, and Br atoms, respectively. With the ``light'' (``tight'') settings, there is a negligible error of around 60 (5) $\mu$eV/atom between the all-electron and the frozen core results including the pseudo Hamiltonian, the potential energy, and the total energy, suggesting that FC99+C+V produces sufficiently accurate forces for molecular dynamics simulations.

\begin{figure}
\includegraphics[width=0.35\textwidth]{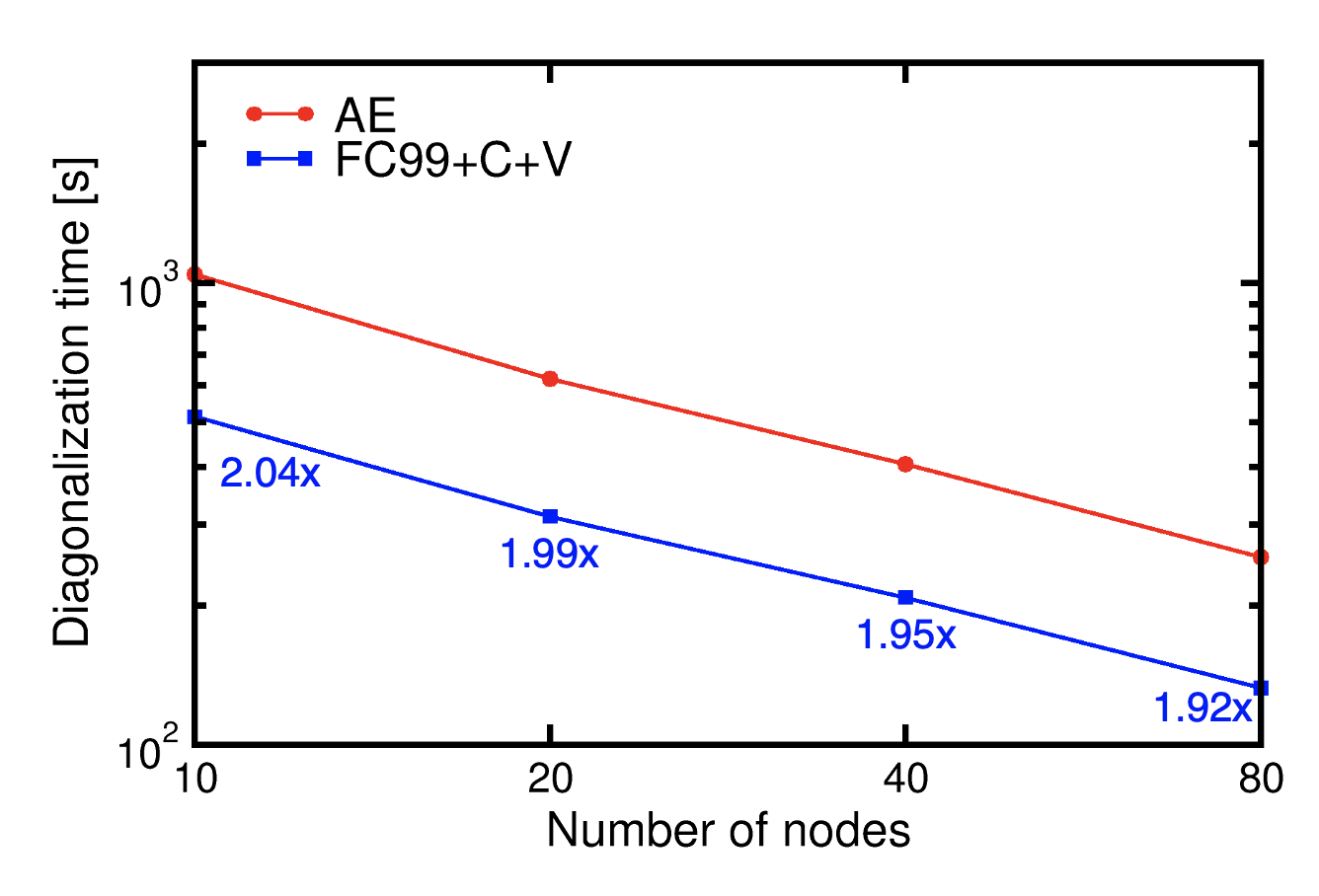}
\caption{Time to solve the generalized eigenproblem in equation~\ref{eq:gevp} for a $4 \times 4 \times 4$ supercell of the CsPbBr$_3$ model in figure~\ref{fig:cspbbr3} (2,560 atoms). The calculations use the ``light'' settings of FHI-aims and a $1 \times 1 \times 1$ $\boldsymbol{k}$-grid. $N_\text{basis} = 94,208$. Red circles: All-electron. Blue squares: FC99+C+V. Orbitals up to the 4s, 4d, and 3s states are frozen for the Cs, Pb, and Br atoms, respectively. $N_\text{core} = 28,672$. The speedup factors due to the frozen core approximation are labeled in the figure.}
\label{fig:time}
\end{figure}

Finally, we report the performance of the FC99+C+V method for large systems. KS-DFT calculations are carried out for a $4 \times 4 \times 4$ supercell of the CsPbBr$_3$ model in figure~\ref{fig:cspbbr3}, with 2,560 atoms in total. The calculations use the PBE functional, the $\Gamma$ point of the reciprocal space, and the ``light'' settings of FHI-aims. The number of basis functions is 94,208. In the frozen core case, orbitals up to the 4s, 4d, and 3s states are frozen for the Cs, Pb, and Br atoms, respectively, amounting to 28,672 (30.4\%) core orbitals. Figure~\ref{fig:time} compares the time to solve equation~\ref{eq:gevp} with all-electron and with FC99+C+V. Using 10 to 80 Haswell nodes of the Cori supercomputer, FC99+C+V accelerates the solution of equation~\ref{eq:gevp} by 1.92x $\sim$ 2.04x. The overall speedup for a complete SCF cycle ranges from 1.71x to 1.75x. Clearly, the additional steps in the FC99+C+V implementation scale as well as the ELPA eigensolver for the all-electron case with the number of available CPU cores. Freezing more orbitals is expected to deliver a larger speedup as long as a slightly increased error can be tolerated (see table~\ref{tab:time}).

\section{Conclusions}
\label{sec:summary}
The frozen core approximation is a long existing technique to reduce the computational complexity of electronic structure theory. In this paper, we have presented a highly accurate frozen core approximation, FC99+C+V, to be applied to the diagonalization stage of all-electron KS-DFT using localized basis functions. Our method takes into account the finite precision of the numerical integration in real space. Our results suggest that the proposed FC99+C+V method reliably reproduces the corresponding all-electron results.
\begin{itemize}
\item The FC99+C+V eigenvalues of the Kohn-Sham orbitals agree well with all-electron case. The Kohn-Sham orbitals are properly orthonormalized, yielding sufficiently accurate electron density and robust SCF convergence.

\item The total energies are in excellent agreement with the all-electron results for a benchmark set of 103 materials, with a negligible difference generally below one meV/atom for a core levels defined as lying below $E = -200$ eV. This is, in fact, an important result in its own right since, to the authors' knowledge, the precision of the frozen core approximation without any other intervening factors has not been established across a benchmark set of this breadth before.

\item The atomic forces are sufficiently accurate to drive molecular dynamics simulations.
\end{itemize}

KS-DFT calculations of CsPbBr$_3$ models of various sizes (up to 2,560 atoms) show that the FC99+C+V method, by reducing the dimension of the eigenproblem, are considerably faster than regular all-electron calculations for systems made of heavy elements. The efficiency and accuracy of the method can be systematically controlled by tuning the number of frozen orbitals. The total energy error when freezing all orbitals below $E = -200$ eV is practically negligible. Practically the same core orbitals as in their all-electron counterpart are retained in the full potential (no shape approximations made). Finally, the computational cost analysis shown here is still restricted to the eigensolver only, while further computational savings should be straightforward to realize in other computationally expensive steps (notably, the update of the electron density and its gradients on a real-space grid). The FC99, FC99+C, and FC99+C+V methods are implemented in the Electronic Structure Infrastructure (ELSI) software~\cite{elsi_yu_2018,elsi_yu_2020} with a generic interface that can be adopted by any all-electron electronic structure codes employing localized basis sets, provided that accurate atomic orbitals for the core electrons are available from free atom calculations.

\section*{Data Availability Statement}
The data that support the findings of this study are openly available online~\cite{raw_data}.

\section*{Supplementary Material}
See supplementary material for a list of frozen orbitals of the elements in the 103-material benchmark set, and total energies and total energy errors of the benchmark systems presented in this paper.

\begin{acknowledgments}
This research was supported by the National Science Foundation (NSF) under Award No. 1450280. Yu was supported by a fellowship from the Molecular Sciences Software Institute under NSF Award No. 1547580. This research used resources of the National Energy Research Scientific Computing Center (NERSC), a U.S. Department of Energy Office of Science User Facility operated under Contract No. DE-AC02-05CH11231. We thank the anonymous reviewers of this paper for their constructive feedback. In particular, the description of the frozen core approximation in section~\ref{sec:fc99} was adapted from an accurate and concise summary provided by one of the reviewers.
\end{acknowledgments}

\nocite{*}
\bibliographystyle{apsrev4-1}
\bibliography{aipsamp}

\end{document}